\documentclass[aps,prl,twocolumn,a4paper]{revtex4}
\usepackage{amsmath}
\usepackage{amsfonts}
\usepackage{graphicx}
\usepackage{pstricks}
\usepackage{bbm}

\def\CC{{\rm\kern.24em \vrule width.04em height1.46ex depth-.07ex
\kern-.30em C}}
\newcommand{\expect}[1]{\left\langle #1 \right\rangle}
\def\trace{{\rm tr}\;}
\newcommand{\bigfrac}[2]{\mbox {${\displaystyle \frac{ #1 }{ #2 }}$}}
\newcommand{\beq}{\begin{equation}}
\newcommand{\beqa}{\begin{eqnarray}}
\newcommand{\eeq}{\end{equation}}
\newcommand{\eeqa}{\end{eqnarray}}

\newcommand{\ket}[1]{\left | #1 \right\rangle}

\begin{document}

\title{Enhancement of pairwise entanglement via $\mathbbm{Z}_2$ symmetry breaking}

\author{Andreas Osterloh}
\author{Guillaume Palacios}
\affiliation{Institut f\"ur Theoretische Physik, Leibniz Universit\"at Hannover,
             Appelstr.\ 2, 30167 Hannover, Germany}

\author{Simone Montangero}
\affiliation{NEST-CNR-INFM \& Scuola Normale Superiore, Piazza Cavalieri 7, I-56126
  Pisa, Italy}
\affiliation{Institut f\"ur Theoretische Festk\"orperphysik, Universit\"at Karlsruhe,
    76128 Germany}

\date{\today}

\begin{abstract}
We study the effect of symmetry breaking in a quantum phase transition
on pairwise entanglement in spin-1/2 models. We give a set of 
conditions on correlation functions a model has to meet in order to
keep the pairwise entanglement unchanged by a parity symmetry
breaking. It turns out that all mean-field solvable models do 
meet this requirement, whereas the presence of strong correlations leads to a
violation of this condition. This results in an order-induced enhancement of
entanglement, and we report on two examples where this takes place.
\end{abstract}
\maketitle
Entanglement is a non-locality inherent to quantum mechanics and an 
important resource for quantum optics and quantum information processing. 
It also has attained
a lot of interest in the last decade from condensed matter
physicists, especially since a connection between quantum non-locality
and quantum phase transitions was proposed~\cite{OstNAT,Osborne02}.
This initiated a vast analysis of quantum critical models with respect
to their entanglement features based on the few computable
entanglement measures at hand~\cite{VidPalMos,Popp04,Verstraete04,
Laflorencie05,Oliveira06,Larsson05,Its,Peschel,GuLiLin04,Amico05,Amico06,
GuTianLin04,Firenze04,Firenze05,Alcaraz04}. 
One of the entanglement measures probed on such models
is the concurrence~\cite{Hill97,Wootters98,Uhlmann00}, 
a measure of pairwise entanglement for spins-1/2 (qubits).
Most of the results obtained in the literature make use of
certain symmetries of the model Hamiltonian which
reflect in the reduced density matrix.  
In presence of quantum phase transitions, a symmetry is typically broken 
in the {\em ordered} region with non-vanishing order parameter.
The effect of the broken symmetry on the
entanglement largely has not been taken into account in the existing
literature (but see e.g.~\cite{Osborne02} for the local entropy).  It
was exactly this subject which has been addressed by Sylju{\aa}sen
in~\cite{Syl,Syl2} for the quantum-Ising and the ferromagnetic XXZ
model, where a parity symmetry is broken at the transition. He
provided a condition on correlation functions that guarantees
the invariance of the concurrence under symmetry
breaking. The models under consideration there fulfill this condition.
The focus of the present work is to investigate the regime where 
Sylju{\aa}sen's condition is violated as it happens for the 
XY model in transverse field. 

In this letter, we first summarize the main result
of Ref.~\cite{Syl}. Then, we derive the condition for two-point
correlation functions and form factors that guarantees the invariance
of the concurrence under parity symmetry breaking where the Sylju{\aa}sen condition does not apply. Finally, we illustrate our result, showing that mean-field
solvable models satisfy this condition and demonstrating that symmetry breaking in general must be given account for, both for the XY spin chain in
transverse magnetic field and the Lipkin-Meshkov-Glick (LMG) model.

We will next review the main result of Ref.~\onlinecite{Syl}, 
adopting for the sake
of cross-readability the notation used in this work.
The model Hamiltonians $H$ under consideration have a parity symmetry or 
global phase flip symmetry, meaning that the eigenstates can be chosen as
superpositions from states all having the same parity of flipped spins 
(i.e.: odd or even). It is reflected by $[\prod_i\sigma_i^z,H]=0$. 
The general 2 sites density matrix of such a system is given by
\begin{equation}
 \label{Isingdensitymatrix}
 \rho_{ij} =\left(
 \begin{array}{llll}
    A & a & a & F \\
    a & B & C & b \\
    a & C & B & b \\
    F & b & b & D
\end{array}
\right),
\end{equation}  
where indices for the entries of the density matrix have been omitted.
The entries of $\rho$ are related to spin correlators as follows:
$a=(\expect{ S^x }+ 2\expect{ S^x S^z })/2$,
$b=(\expect{ S^x } - 2\expect{ S^x S^z })/2$,
$A=\expect{ S^z S^z } + \expect{ S^z } +1/4$,
$B=1/4 - \expect{ S^z S^z }$,
$C=\expect{ S^x S^x } + \expect{ S^y S^y }$,
$D=\expect{ S^z S^z } - \expect{ S^z } +1/4$ and
$F=\expect{ S^x S^x } - \expect{ S^y S^y }$.
The symmetry-breaking manifests itself in $a,b\neq 0$. 
For $a=b=0$, the square root of the eigenvalues of 
$\rho \tilde{\rho}$ are
$B \pm C$ and $\sqrt{AD} \pm F$. The concurrence of a 2-site reduced
density matrix $\rho$ is computed from the 
positive semidefinite matrix 
$R:=\rho ( \sigma_y\otimes\sigma_y )\rho^{*} ( \sigma_y\otimes\sigma_y )$ as
${\cal C}=\max\{0,2\lambda_{max}-\trace \sqrt{R}\}$, where $\lambda_{max}$ is 
the maximum eigenvalue of $\sqrt{R}$. Thus the concurrence is
${\cal C} = 2 \max \left\{ 0,|C|-\sqrt{AD}, |F|-B \right\}=:
2\max\left\{ 0,{\cal C}_{af},{\cal C}_f\right\}$.
We refer to the concurrence as belonging to 
the anti-ferro- or ferromagnetic sector whenever the largest eigenvalue 
of $\sqrt{R}$ is from this sector and denote it by 
${\cal C}_{af}$ and ${\cal C}_f$, respectively.
Throughout all the letter will we adopt the notation
$\expect{ S^\alpha S^\beta }:= \expect{ S^\alpha_i S^\beta_j }$
and $\expect{ S^z }:=(\expect{ S^z_i }+\expect{ S^z_j })/2$.
For the models considered here, the Hamiltonian is real
and so is the density matrix.

In order to study the effect of symmetry breaking, accompanied by non-zero 
$\expect{ S^x S^z }$ and order parameter $\expect{ S^x }$,
one has to extract the dependence of the eigenvalues of $R$ on $a$ and $b$.
This task simplifies since the eigenvector $(0,1,-1,0)$ 
(the singlet Bell state) and its eigenvalue $B-C$ are unchanged. 
If $C>0$, this robust eigenvalue can not become the largest one.

The exact factoring of a cubic polynomial is not very handy and 
has been cleverly avoided by Sylju{\aa}sen, when concentrating on the
coefficients of the characteristic polynomial 
$X^3-g_2 X^2+g_1 X - g_0$ of $R$ with 
$g_0=(xyz)^2$, $g_1 = x^2 y^2 + x^2 z^2 + y^2 z^2$, and
$g_2 = x^2 + y^2 + z^2$
rather than studying the eigenvalues of $\sqrt{R}$ themselves. 
The connection between the $g_i$'s and the entries of the reduced 
two-spin density matrix is~\cite{Syl}
\begin{eqnarray}
g_0 & = & (\alpha^2-4\gamma \delta) \beta -4\mu \nu \alpha -4\mu^2
  \delta -4\nu^2 \gamma \\ g_1 & = & \alpha^2 +2\alpha \beta -4\mu \nu
  -4\gamma \delta \\ g_2 & = & 2\alpha + \beta
\end{eqnarray}
where $\alpha = F^2+AD-2ab$, $\beta = (B+C)^2-4ab$, $\gamma =DF-b^2$, 
$\delta=AF-a^2$, $\mu=aD-b(B+C-F)$ and $\nu=a(B+C-F)-bA$.
The concurrence would then simply be expressed
as ${\cal C}=\max\{0,\kappa-(B-C)\}$, with $\kappa:=z-x-y$ and $z$ the 
largest eigenvalue of $\sqrt{R}$. 
The condition that the concurrence be unaffected by symmetry breaking
is then~\cite{Syl}
\beq\label{CConstC+}
2\kappa\sqrt{g_0}=\left(\bigfrac{\kappa^2-g_2}{2}\right)^2-g_1\; .
\eeq
Interestingly, for dominating ${\cal C}_f$, i.e.
when $\sqrt{AD}+|F|$ is largest, this condition is identically satisfied which is the main outcome of Ref.~\cite{Syl}.  

We will now extend the analysis of Ref.~\cite{Syl} to
the regime where ${\cal C}_{af}$ dominates, but still $C>0$.
The case $C<0$ and $B-C$ being the largest eigenvalue of $\sqrt{R}$, 
is treated the same way, just replacing $\kappa$ by
$\tilde{\kappa}:=x+y+z$.

It is a generic feature of quantum spin models
(with competing interactions) 
that there is a point in coupling parameter space at which the 
ground state factorizes~\cite{Barouch71,Kurmann,Firenze04}.
At such a {\em factorizing field} any measure of 
entanglement 
must vanish. 
If at both sides of a factorizing point
the concurrence is non-zero, there are two scenarii: either the concurrence 
is smooth at $h_f$ or the sector giving support to the 
concurrence necessarily has to change at this point.
It is the latter scenario which actually takes place in the models examined
so far in the literature as e.g. 
the transverse XY model~\cite{OstNAT,OsterlohSPIE},
the XYZ model~\cite{Firenze04,Firenze05}, 
and the Lipkin-Meshkov-Glick model (LMG)~\cite{VidPalMos,DusVidPRB};
the Sylju{\aa}sen condition applies between the factorizing and
the quantum critical field. 
The explicit form of the concurrence for parity symmetric models
and translational symmetry 
(which includes all above cited studies) 
implies that ${\cal C}_{af}={\cal C}_f=0$
at the factorizing field, and they cross 
unless $2\langle S^y S^y\rangle - (\langle S^zS^z\rangle- \langle
S^z \rangle^2)/(2\langle S^x\rangle^2)\sim(h-h_f)^n$ 
with even $n>0$ (for symmetry breaking field in x-direction).
If only one crossing occurs, then the concurrence
is robust against $\mathbbm{Z}_2$ symmetry breaking between critical and 
factorizing point.
Below the factorizing field ${\cal C}_{af}$ dominates
and the concurrence will in general be affected by the broken symmetry.
Nonetheless, to the best of our knowledge 
no such case has been reported in the literature, which
motivates the search for
further conditions that make the concurrence robust
against symmetry breaking or for examples
where the concurrence {\em is} affected. We will show in this letter that both
cases occur.

Having taken the square of Eq.(\ref{CConstC+}) and
inserted $\kappa\rightarrow B+C-2\sqrt{AD}$, 
the resulting expression is conveniently expressed in the new 
variables $\lambda:=b/a$ and $a$, leading to
\beq
\label{NewCondC+} 
32a^2(\lambda-\lambda_0)^2\left[1+a^2(\lambda-\lambda_1)(\lambda-\lambda_2)\right]
=0\quad , 
\eeq 
where 
$\lambda_0 = \sqrt{D/A}$, $\lambda_1+\lambda_2=2\kappa/A$ and $(\lambda_1-\lambda_2)^2 =4\kappa(B+C)/A^2$. 
 
Both $\lambda_1$ and $\lambda_2$ are real for $\kappa \ge 0$,  which is 
mandatory for a non-vanishing concurrence.
The real solutions to Eq. (\ref{NewCondC+}) are constraints on the 
correlation functions of the model which insure the concurrence to be 
unaffected by symmetry breaking. 
The solution $\lambda=\lambda_0$ can then be recast into a simple condition: 
\beq
\label{cond0} 
\frac{\sqrt{(1+4\expect{ S^zS^z })^2-16 \expect{ S^z }^2}}
{1+4\expect{ S^z } + 4\expect{ S^z S^z }} \equiv \frac{\expect{ S^x }
- 2 \expect{ S^x S^z}}{\expect{ S^x } +2 \expect{ S^x S^z }} \; .
\eeq 
Note that this condition is automatically satisfied for mean-field 
solvable models as a direct consequence of the factorization of the 
two-point functions in this case.

Another real solution to Eq.(\ref{NewCondC+}) exists if $a\le -1$ and
\beq
\label{csx}
\expect{S^x}\left(\bigfrac{\lambda_1-\lambda_2}{\lambda_1+\lambda_2+2}\right)^2=
\pm\frac{1}{\pi}
\eeq 
to insure the reality of both $a$ and $b$.
This leads to the following transcendental equation
\beq
\kappa \frac{B+C}{(\kappa +A)^2}\ln\left(\frac{\kappa}{A}|a|-1\right)=
\frac{A(\kappa + A)}{A(\kappa + A)+2\kappa^2}\frac{1}{\pi|a|}\; .
\label{ta}
\eeq
Eq. (\ref{ta}) has a unique real solution for arbitrary $A$, $B$ and $\kappa$. 
The rewriting of this condition in terms of correlation function is lengthy 
but straightforward.
There is numerical evidence that $a < -1$ only occurs for 
a two-site reduced density matrix of rank $3$ or less.
Interestingly, rank 3 and 2 are assumed for 
the (2-magnon) W state $\sum_i w_{ij} S^+_iS^+_j\ket{\Downarrow}$
and GHZ (cat) state $\ket{\Downarrow}+\ket{\Uparrow}$ of the whole chain, 
where $w_{ij}\in\CC$ and $\ket{\Uparrow}$ and $\ket{\Downarrow}$ are 
in opposite direction fully polarized states.
If Eqs. (\ref{cond0}) and (\ref{csx})--(\ref{ta}) are violated, the
concurrence will be affected by symmetry-breaking, as 
condition (\ref{CConstC+}) is then violated. 

\underline{LMG model}~~
The LMG model~\cite{LMG1} in transverse field is described by the Hamiltonian
\begin{equation}
\label{Hlmg}
H_{LMG}=-\frac{1}{N}\sum_{i<j}
\left( \sigma_{x}^{i}\sigma_{x}^{j} +\gamma\sigma_{y}^{i}\sigma_{y}^{j} \right)
      -h \sum_{i}\sigma_{z}^{i}\end{equation}
where the $\sigma_{\alpha}^i$ are the Pauli matrices for the $i$-th
lattice site and $N$ is the chain length. 
Any two spins are interacting with the same coupling strength. 
The prefactor $1/N$ leads to a finite free energy per spin in the
thermodynamic limit. $H$ commutes with the total spin and the 
spin-flip operator $\prod_i \sigma_z^i$ for any 
anisotropy parameter $\gamma$. 
This discrete $\mathbbm{Z}_2$ symmetry is broken at the quantum
critical point, $h_c=1$. The factorizing field of this model is
$h_f=\sqrt{\gamma}$. Due to its infinite-range coupling, the
mean-field approximation for the LMG model becomes exact in the
thermodynamic limit~\cite{Botet83}.
In this limit, condition (\ref{cond0}) is
satisfied for arbitrary transverse field. 
The non-trivial behavior of the entanglement of the
LMG lies in the finite-size corrections to the mean-field
solution~\cite{VidPalMos,DusVidPRB}. 
\begin{figure}[ht]
\begin{center}
\includegraphics[width=0.42\textwidth]{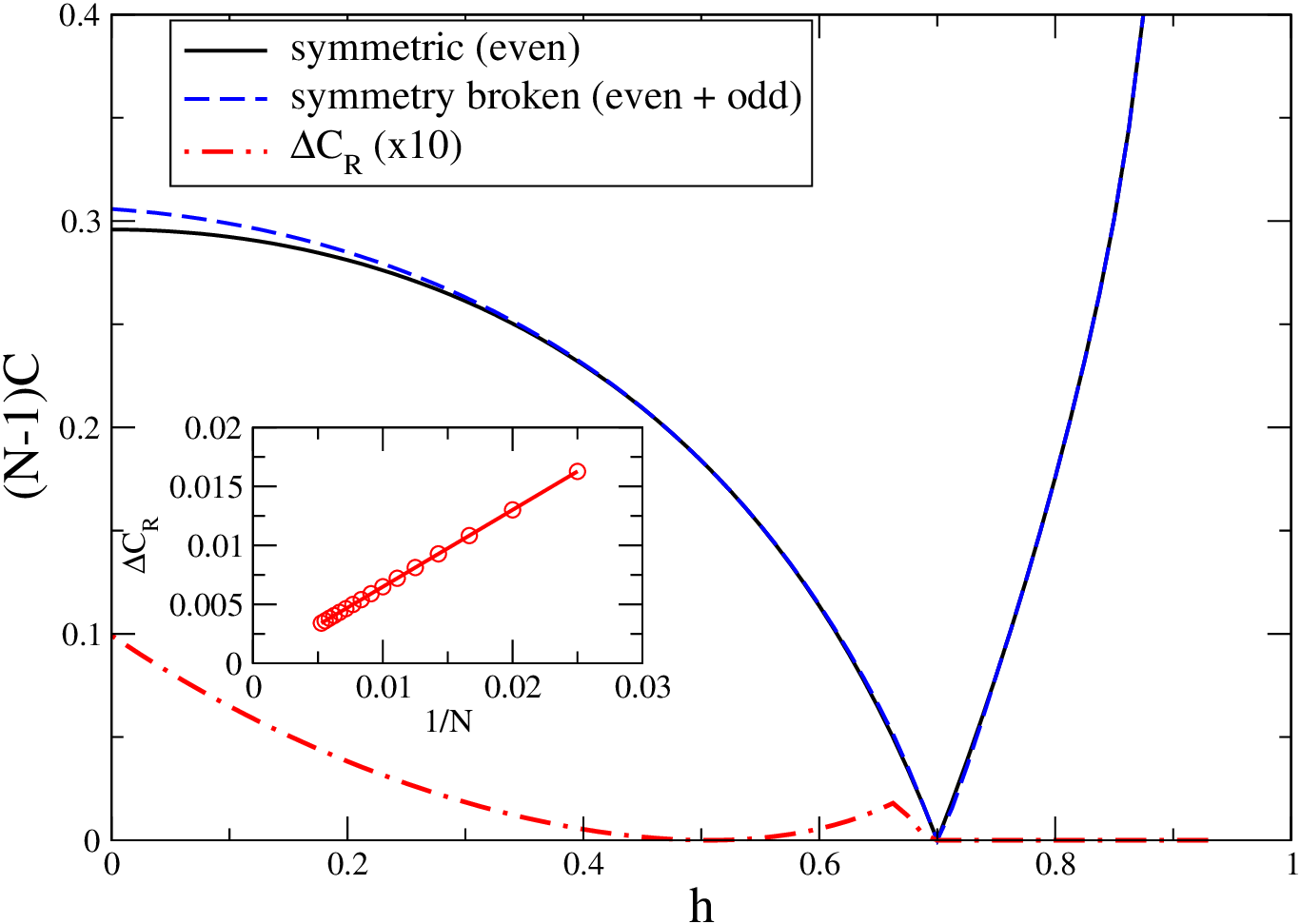}
\caption{
\label{Clmg}
Rescaled concurrence for the LMG model: The even sector (full line) 
and the symmetry broken (dashed line)
concurrence is shown for $N=100$ and $\gamma=0.5$; the inset shows the
difference between the rescaled concurrence in the
$\mathbbm{Z}_2$-symmetric case and in the $\mathbbm{Z}_2$-broken case for
fixed magnetic field $h=0.2<h_f$. It scales as 
$a_0 /N + a_1$, with $a_0=0.651622$ and $a_1 \simeq 10^{-5}$. 
The dash-dotted curve is this difference rescaled by a factor of 10
as a function of the magnetic field. It goes to zero at $h\sim 0.5$.
At such a point, one of the conditions for the correlation
functions of the model must be satisfied.}
\end{center}
\end{figure}

In Fig. \ref{Clmg} we report numerical data for the
rescaled concurrence, ${\cal C}_R = (N-1) {\cal C}$, for the even (black full curve)
and $\mathbbm{Z}_2$-broken (blue dashed curve) ground state. 
In the thermodynamic limit
($N$ goes to $\infty$ while the free energy per spin remains finite),
both curves coincide. 
The analytic expression of this limiting curve in the region $0
\leq h \leq h_f$ has been obtained in~\cite{DusVidPRB}:
\begin{equation}
\lim_{N\to\infty} {\cal C}_R^{Sym} = 1 - \sqrt{\frac{1-\gamma}{1-h^2}}\; .
\end{equation}
For finite $N$, we clearly see a deviation of the
concurrence calculated with a state which violates the $\mathbbm{Z}_2$
spin-flip invariance from the concurrence evaluated with a state which
conserves the latter symmetry. This is indeed a finite-size effect
which is beyond the scope of the first order quantum corrections performed
in Ref.~\cite{DusVidPRB}. In fact, the difference between ${\cal
C}^{Sym}$ and ${\cal C}^{Broken}$ is of order $1/N^2$ as shown in 
the inset of Fig. \ref{Clmg} \footnote{an analytical proof of this assumption is left to a future publication.}.

\begin{figure}[ht]
\begin{center}
\includegraphics[width=0.42\textwidth]{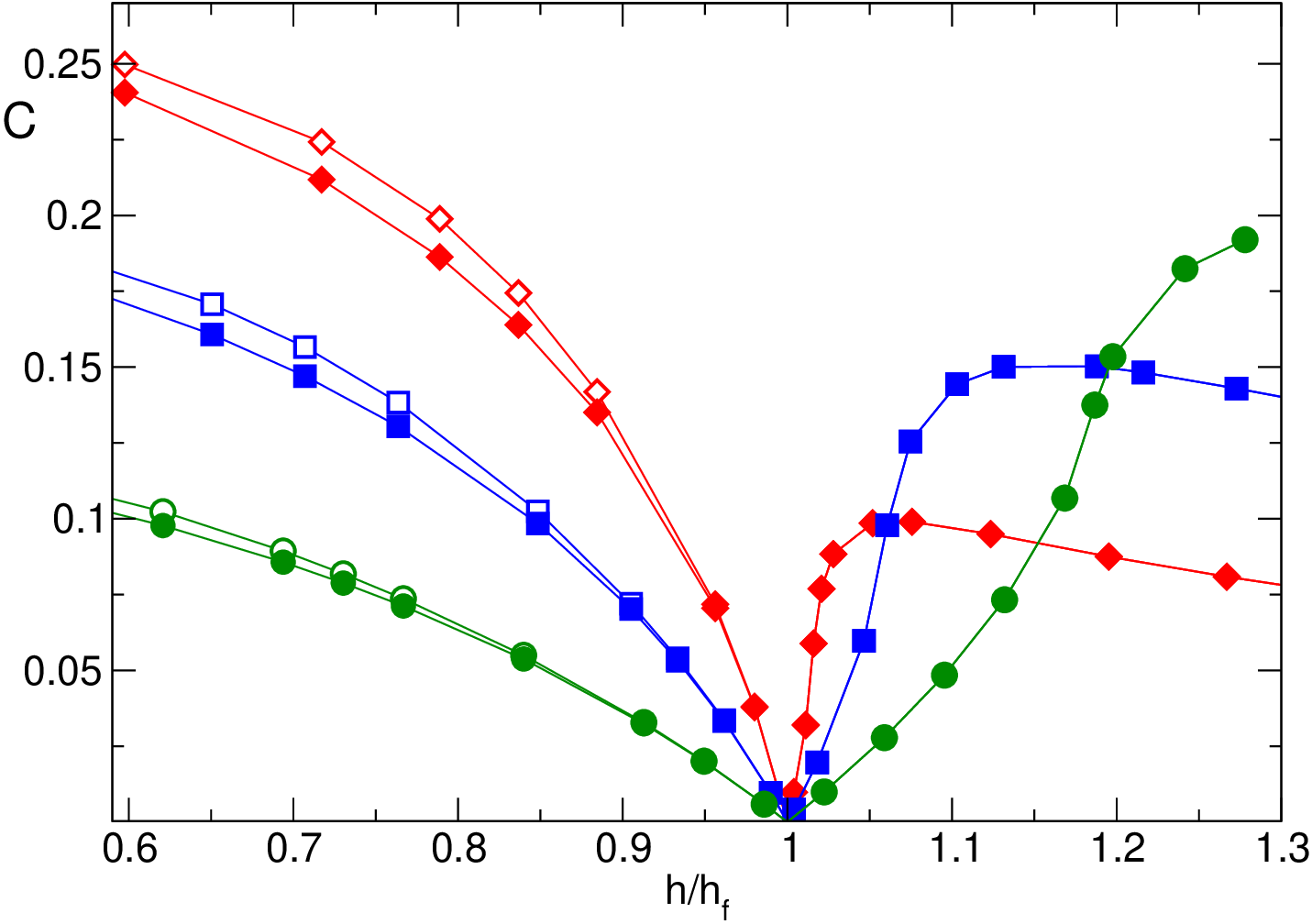}
\caption{
\label{CXY}
N.n. concurrence for the anisotropic XY model for
$N=199$, $\gamma=0.3,0.5,0.7$ (circles, squares and diamonds
respectively), $m=200$
in the even sector (full) and for the state with $h_B=10^{-6}$ (empty).} 
\end{center}
\end{figure}

\begin{figure}[ht]
\begin{center}
\includegraphics[width=0.42\textwidth]{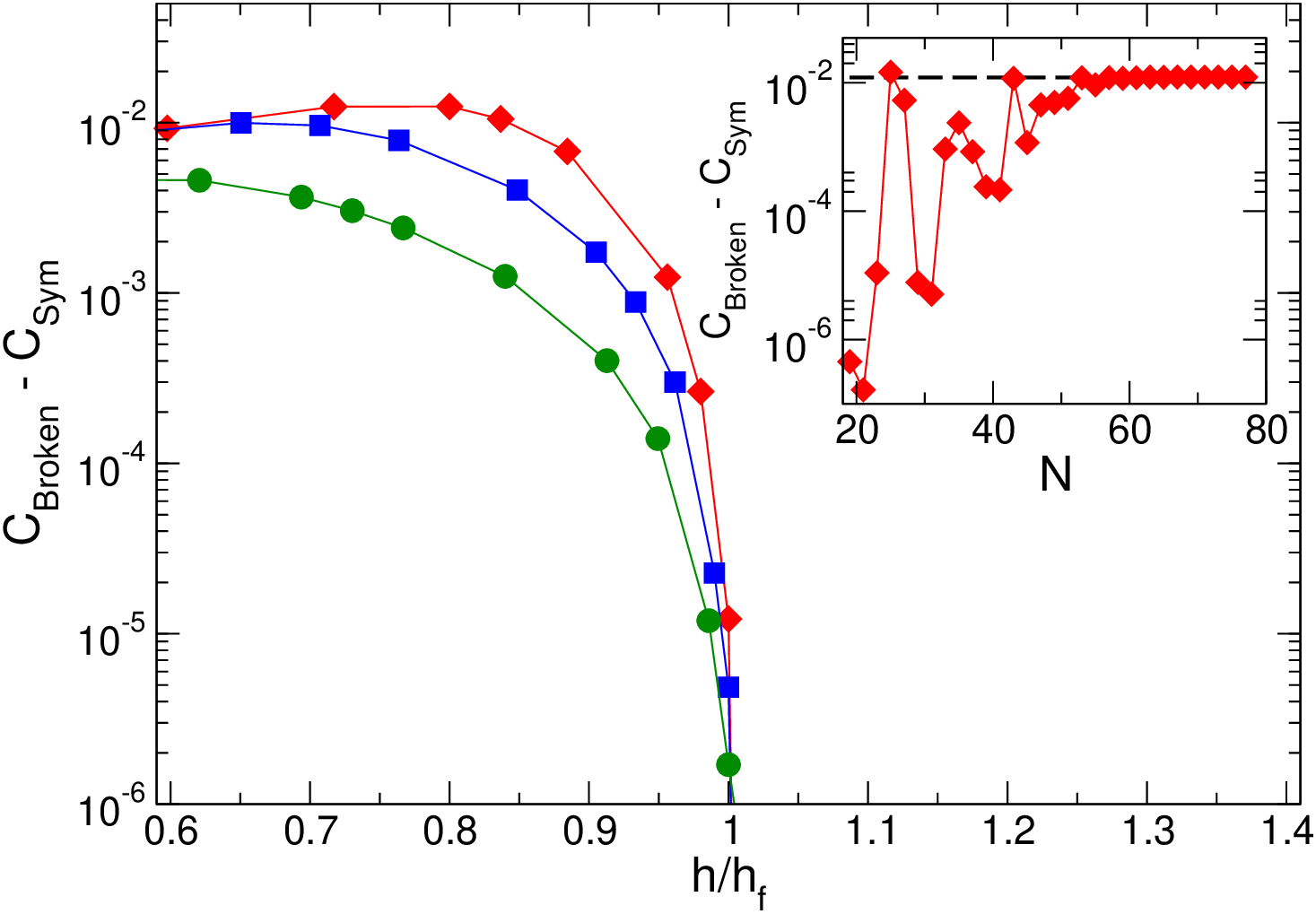}
\caption{
\label{diffCXY}
Difference of the n.n. concurrence in the even sector and for the 
symmetry broken ground state from
the data of Fig.~\ref{CXY} as a function of the transverse field. 
The maximum relative deviation amounts to around $10\%$, decreasing
with $\gamma$ for sufficiently small $h$.
Inset: Finite size scaling for $\gamma=0.7$ and $h/h_f=0.8$ (diamonds) 
and limiting value (dashed line).}
\end{center}
\end{figure}

\underline{XY model}~~
The transverse XY model is defined by the Hamiltonian~\cite{Barouch71}
\begin{equation}
\label{HXY}
H_{XY}=- \frac{1}{4} \sum_{i}
\left( \sigma_{x}^{i}\sigma_{x}^{i+1} + \gamma \, \sigma_{y}^{i}\sigma_{y}^{i+1} \right)
      -h \sum_{i}\sigma_{z}^{i},
\end{equation}
where $h$ is the transverse field strength and $\gamma$ the anisotropy
parameter. Below the quantum critical field
$h_c=(1+\gamma)/4$ the parity symmetry is broken;
the factorizing field is at $h_f=\sqrt{\gamma}/2$. 
We numerically computed the ground state nearest neighbor (n.n.) 
concurrence by means of the
Density Matrix Renormalization Group~\cite{dmrgrev}. 
This powerful numerical technique finds an optimal truncated bases of size
$m\sim 100-200 \ll 2^N$ to describe the spin chain wave function keeping 
the desired precision~\cite{dummies}. 
In Fig.~\ref{CXY} we show the results of the numerical simulations for
the concurrence of the central spins (in order to minimize boundary effects) 
of a chain of length $N=199$ and different anisotropy values. 
The concurrence for the even sector  ${\cal
C}^{Sym}$ and broken symmetry ground state ${\cal C}^{Broken}$ 
are then compared for different anisotropy values. 
The concurrence vanishes at the factorizing
field and its derivative with respect to the transverse
field at the critical point diverges as expected~\cite{OstNAT}. 
The even and odd
sector concurrences are found to coincide for the parameter range we 
considered (data not shown).
We used an additional field $h_B$ along $\sigma_x$ to break the symmetry; 
our results are stable with respect to changes in the field strength
$h_B \in [10^{-4}:10^{-8}]$ and the field direction in the $XY$ plane
(data not shown).
Below the factorizing field the concurrence of the broken symmetry and even
ground state are clearly different while they are equal between
the factorizing and the critical field as there the condition (\ref{CConstC+}) 
holds~\cite{OsterlohSPIE}. 
In Fig.~\ref{diffCXY} we plot the difference between the broken
symmetry concurrence and the concurrence in the even sector from the
data of Fig.~\ref{CXY}. As expected, for $h/h_f \ge  1$ the difference
is zero while on the left of the factorizing field ($h/h_f <  1$) the
difference is not negligible. The inset of Fig.~\ref{diffCXY} reports
the finite size scaling of $|{\cal C}^{Sym}$ - ${\cal
 C}^{Broken}|$. Differently from the LMG Model, the broken
symmetry effect on the concurrence does not vanish in the
thermodynamic limit.
It approaches a non-zero value with oscillating behavior.
This means that for the $XY$ model
conditions~(\ref{cond0})-(\ref{ta}) and 
thus Eq.~(\ref{CConstC+}) 
are violated below $h_f$. 

\underline{Conclusions}~~We have found 
conditions on spin correlation functions, 
which ensure that the 
concurrence is invariant respect to breaking of a parity symmetry.
They are necessary and sufficient in a regime complementary to where
the relations from Ref.~\cite{Syl} do apply.
For mean field exact models, one of the conditions (Eq.~\ref{cond0})
is satisfied and hence will the rescaled concurrence be unaffected by
parity symmetry breaking.
The further conditions (Eqs.~\ref{csx},~\ref{ta}) only emerged for
the reduced two-site density matrix having rank three or less.
Interestingly, this occurs for the whole system being in  
a W-type state with two 
running flipped spins in the ferromagnetically polarized state
and a W or GHZ-state, respectively. 
It would be interesting to verify, 
whether such states may satisfy the corresponding condition
on correlation functions; W-states would lead to a long range 
concurrence, which has been observed in~\cite{Amico06} close to the 
factorizing point. 
Numerical studies reveal that the concurrence is affected 
by symmetry breaking for the LMG and the XY model.
As a consequence, the conditions (\ref{cond0})-(\ref{ta})
on the spin correlations are violated up to perhaps a single value 
for the magnetic field in the LMG model.
We gave certain conditions that might admit
scenarii different from a robust concurrence between $h_c$
and $h_f$.
Their investigation could provide important insight in
the interplay of entanglement and quantum phase transitions.

Fruitful discussions with R. Fazio and
financial support from EUROSQIP, CRM ``E.
De Giorgi'' of SNS, the Alexander Von Humboldt Foundation (SM)
and the Graduiertenkolleg 282
(GP) are acknowledged.
The DMRG code released
within the ``Powder with Power" project (www.qti.sns.it)
has been used.

\end{document}